# Enhanced Two-Photon Absorption Using Entangled States and Small Mode Volumes

Hao You, S.M. Hendrickson, and J.D. Franson
*University of Maryland, Baltimore County, Baltimore, MD 21250*

We calculate the rate of two-photon absorption for frequency-entangled photons in a tapered optical fiber whose diameter is comparable to the wavelength of the light. The confinement of the electric field in the transverse direction increases the intensity associated with a single photon, while the two-photon absorption rate is further enhanced by the fact that the sum of the frequencies of the two photons is on resonance with the upper atomic state, even though each photon has a relatively broad linewidth. As a result, the photons are effectively confined in all three dimensions and the two-photon absorption rate for frequency-entangled photons in a tapered fiber was found to be comparable to that for unentangled photons in a microcavity with a small mode volume.

## I. INTRODUCTION

In addition to its fundamental interest, two-photon absorption has recently been shown to have potential applications in quantum information processing and classical communications. Quantum logic operations can be performed using the quantum Zeno effect produced by strong two-photon absorption [1, 2], while the resolution of images can be enhanced using two-photon absorption in conjunction with quantum imaging techniques [3]. Two-photon absorption can also be used to implement a source of single photons [4], and it was recently shown that all-optical switching of classical beams of light can be performed at very low power levels using two-photon absorption [5]. All of these potential applications would benefit from new techniques for enhancing the rate of two-photon absorption while minimizing single-photon losses.

In an earlier paper, we showed that two-photon absorption can be enhanced in tapered optical fibers with diameters less than the free-space wavelength of the photons [6]. A substantial fraction of the energy propagates outside of such a tapered fiber, which allows the photons to interact with an atomic vapor. The reduction in the effective mode volume compresses the photons into a smaller region of space and increases the electric field strength associated with a single photon. Since the rate of two-photon absorption depends on the fourth power of the electric field, this enhances the rate of two-photon absorption compared to the single-photon losses.

Here we generalize this situation to include photons that are entangled in energy and time. Although both photons have a relatively wide bandwidth, the sum of their frequencies is still well-defined, which allows two-photon absorption to be on resonance with the upper excited level of the atoms. Pairs of energy-time entangled photons are emitted at very nearly the same time and will be detected at nearly the same location. This allows both photons to simultaneously interact with a given atom, which enhances the rate of two-photon absorption [7-13]. This enhancement is roughly equivalent to what would occur if the photons were confined to a small region in the direction of propagation, as will be discussed in more detail below. When the entangled photons are also propagating in a tapered fiber, their mode volume is effectively reduced in all three dimensions and the rate of two-photon absorption is greatly increased.

Section II describes the system of interest and the corresponding Hamiltonian. The rate of two-photon absorption (TPA) is calculated in Section III using density-operator techniques, and the rate of two-photon absorption is optimized as a function of the bandwidth and detuning of the photons in Section IV. For comparison purposes, the rate of two-photon absorption in toroidal microcavities with small mode volumes is calculated in Section V for the case of monochromatic (unentangled) photons, and these results are found to be comparable to that obtained for entangled photons in a tapered fiber. A summary and conclusions are presented in Section VI, and further details of the calculations are described in the Appendix.

## II. FREQUENCY-ENTANGLED PHOTONS IN TAPERED FIBERS

The tapered fiber system of interest here is similar to the one studied in Ref. [6], except that here we consider the situation in which two incident photons are frequency-entangled and traveling in the same direction. As illustrated in Fig. 1, a tapered optical fiber with diameter $D$ and length $L$ is surrounded by rubidium vapor with density $\rho_A$. For sufficiently small diameters, only a single transverse mode can propagate in the fieber and the electric field operator can be written in cylindrical coordinates in the form

$$\begin{aligned}\hat{\mathbf{E}}(r,\varphi,z) &= \hat{\mathbf{E}}^{(+)}(r,\varphi,z) + \hat{\mathbf{E}}^{(-)}(r,\varphi,z) \\ &= \sum_\omega N_\beta (E_r \mathbf{e}_r + E_\varphi \mathbf{e}_\varphi + E_z \mathbf{e}_z) e^{i\beta(\omega)z} \hat{a}(\omega) + H.c.\end{aligned} \quad (1)$$

Here $E_r$, $E_\varphi$, and $E_z$ are the classical components of the electric field $\mathbf{E}(r,\varphi,z) = (E_r \mathbf{e}_r + E_\varphi \mathbf{e}_\varphi + E_z \mathbf{e}_z) e^{i\beta(\omega)z}$ as derived by Tong *et al*. [14]. The operator $\hat{a}(\omega)$ annihilates a photon with angular frequency $\omega$ in the fiber, $N_\beta$ is a suitable normalization factor, and *H.c.* denotes the Hermitian conjugate term. The propagation constant $\beta(\omega)$ is a function of the angular frequency $\omega$ and can be determined by solving the appropriate eigenvalue equation [6, 14].

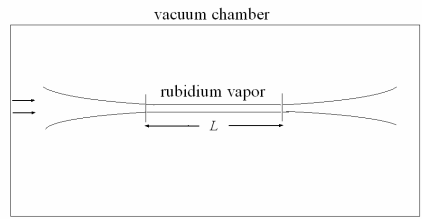

Figure 1. A tapered optical fiber with diameter D in a vacuum chamber surrounded by rubidium vapor. Two frequency-entangled photons are assumed to be propagating in the same direction. The evanescent fields of the photons extend into the atomic vapor and allow strong two-photon absorption to occur.

Suitable pairs of frequency-entangled photons can be produced by spontaneous parametric down-conversion (SPDC) using a $\chi^{(2)}$ nonlinear crystal, for example. (This requires a pump laser with a very narrow bandwidth that is locked to the appropriate three-level transition [15]). As usual, the output of the SPDC crystal will consist of two fields with relatively large sidebands, commonly referred to as the signal and idler. Their angular frequencies $\omega_s$ and $\omega_i$ are anti-correlated and the sum of their frequencies is equal to that of the pump laser. As a result, they can be written as $\omega_s = \omega_{s0} + \nu$ and $\omega_i = \omega_{i0} - \nu$, respectively, where $\omega_{s0}$ and $\omega_{i0}$ are their mean frequencies and $\nu$ is the offset of the two frequencies from their means. For simplicity, the intensities of the two fields are assumed to be equal.

The incident photons are coupled to the three-level rubidium atoms via their evanescent fields that extend outside of the fiber. The relevant rubidium atomic states are illustrated in Fig. 2. The atomic transition frequencies from the ground state $|g\rangle$ to the intermediate excited state $|i\rangle$ and from state $|i\rangle$ to the second excited state $|h\rangle$ will be denoted $\omega_1$ and $\omega_2$, respectively. These transitions are associated with electric dipole moments $\mathbf{d}_1$ and $\mathbf{d}_2$, while the transition $|g\rangle \to |h\rangle$ is forbidden in the dipole approximation. The decay rates from the two excited states will be denoted by $\Gamma_1$ and $\Gamma_2$, respectively. Here we assume that the sum of the frequencies of the down-converted pair is equal to the sum of the atomic transition frequencies, e.g. $\omega_s + \omega_i = \omega_1 + \omega_2$.

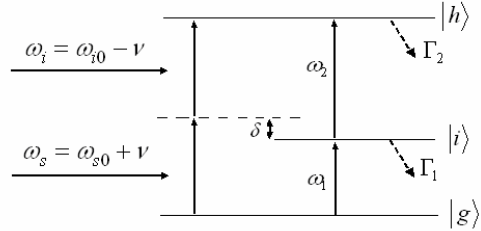

Figure 2. Energy level diagram for the atomic system under consideration. The incident field consists of two entangled photons propagating in the same direction with their angular frequencies centered at $\omega_{s0}$ and $\omega_{i0}$, respectively. The sum of their frequencies is equal to the sum $\omega_1 + \omega_2$ of the atomic transition frequencies from the ground state $|g\rangle$ to the highest highest level $|h\rangle$ through intermediate state $|i\rangle$.

We will first calculate the TPA rate produced by a single atom at a specific location outside of the fiber core, and then later integrate over all possible positions of the atoms, assuming a uniform distribution of atoms at density $\rho_A$. The system of interest will thus consist of a single atom and two incident entangled photons. We assume that all of the atoms are initially in the ground state $|g\rangle$ and that the entangled frequencies of the two photons are described by a Gaussian distribution. (This assumption is reasonable in many down-conversion experiments because of the use of narrow-band frequency filters.) Thus the initial state of the system can be taken to be

$$|\psi_0\rangle = N_\nu \sum_\nu \hat{a}^\dagger(\omega_{s0}+\nu)\hat{a}^\dagger(\omega_{i0}-\nu)e^{-\nu^2/(2\sigma^2)}|0\rangle|g\rangle. \tag{2}$$

Here $\hat{a}^\dagger(\omega_{s0}+\nu)$ and $\hat{a}^\dagger(\omega_{i0}-\nu)$ are the creation operators for the signal and idler modes with frequencies $\omega_{s0}+\nu$ and $\omega_{i0}-\nu$, respectively, $\sigma$ is the half-width of the Gaussian distribution, and $N_\nu$ is a normalization constant to be determined.

For simplicity, we consider the case in which the detuning in the intermediate state $|i\rangle$ is much less for the signal photon than it is for the idler photon, or $|\omega_s - \omega_1| \ll |\omega_i - \omega_1|$. With that assumption, the probability of a two-photon absorption event in which the idler photon is absorbed first followed by the signal photon can be neglected compared to the absorption of the signal photon followed by the idler photon.

In the interaction picture and using the dipole and rotating-wave approximations, the Hamiltonian $\hat{V}$ for the interaction between a single atom and the entangled photons can be written as a superposition of terms with specific values of $\nu$ corresponding to $\omega_s = \omega_{s0} + \nu$ and $\omega_i = \omega_{i0} - \nu$:

$$\hat{V} = \sum_\nu \hat{V}_\nu$$
$$\hat{V}_\nu = m_{1,\nu}^* \hat{\sigma}_{gi}^\dagger \hat{a}(\omega_s) e^{-i\delta t} + m_{2,\nu}^* \hat{\sigma}_{ih}^\dagger \hat{a}(\omega_i) e^{i\delta t} + H.c \tag{3}$$

Here $\hat{\sigma}_{gi}$ produces a transition from $|i\rangle$ to $|g\rangle$, and $\hat{\sigma}_{ih}$ produces a transition from $|h\rangle$ to $|i\rangle$. The atomic matrix elements are $m_{1,\nu}^* = \langle -\mathbf{d}_1 \cdot \mathbf{E}_s(\vec{r})\rangle$ and $m_{2,\nu}^* = \langle -\mathbf{d}_2 \cdot \mathbf{E}_i(\vec{r})\rangle$, where $\mathbf{d}_1$ and $\mathbf{d}_2$ are the corresponding dipole moments. $\mathbf{E}_s(\vec{r})$ and $\mathbf{E}_i(\vec{r})$ are the classical electric fields with propagation constants $\beta(\omega)$ in the tapered fiber as in Eq. (1) [14]. The detuning in the intermediate state is defined as $\delta = \omega_s - \omega_1$.

### III. EQUATIONS OF MOTION AND ANALYSIS METHOD

To calculate the TPA rate, we choose a convenient set of basis states:

$$\begin{aligned}|1_\nu\rangle &= |0\rangle \otimes |h\rangle, \\ |2_\nu\rangle &= \hat{a}^\dagger(\omega_{i0}-\nu)|0\rangle \otimes |i\rangle, \\ |3_\nu\rangle &= \hat{a}^\dagger(\omega_{s0}+\nu)\hat{a}^\dagger(\omega_{i0}-\nu)|0\rangle \otimes |g\rangle.\end{aligned} \tag{4}$$

The time evolution of the density matrix $\hat{\rho}$ is given [6] by

$$\frac{\partial \hat{\rho}}{\partial t} = \frac{-i}{\hbar}[\hat{V}, \hat{\rho}] - \frac{1}{2}\{\hat{\Gamma}, \hat{\rho}\}. \tag{5}$$

The matrix elements of $\hat{V}$ and $\hat{\Gamma}$ will be denoted by $\langle n_v | \hat{V} | m_{v'} \rangle = V_{nv,mv'}$ and $\langle n_v | \hat{\Gamma} | m_{v'} \rangle = \Gamma_{nv,mv'}$. In component form, this equation is equivalent to

$$\frac{\partial \rho_{iv,jv'}}{\partial t} = \frac{-i}{\hbar} \sum_{mv''} \left( (V_{iv,mv''}\rho_{mv'',jv'} - \rho_{iv,mv''}V_{mv'',jv'}) - \frac{1}{2}(\Gamma_{iv,mv''}\rho_{mv'',jv'} + \rho_{iv,mv''}\Gamma_{mv'',jv'}) \right)$$
$$= \frac{-i}{\hbar} \sum_{m} (V_{iv,mv}\rho_{mv,jv'} - \rho_{iv,mv'}V_{mv',jv'}) - \frac{1}{2}(\Gamma_i \rho_{iv,jv'} + \rho_{iv,jv'}\Gamma_j) \tag{6}$$

Here we have used the fact that $V_{nv,mv'} = 0$ unless $v = v'$ and $\Gamma_{iv,mv'} = \Gamma_{mi}\delta_{im}\delta_{vv'}$.

We assume that the effects of any dephasing collisions are negligible and that the system is initially in a pure state $|\psi_0\rangle$. Thus the matrix elements of the initial density matrix can be written as

$$\rho_{iv,jv'}(0) = c_{iv}(0)c^*_{jv'}(0) \tag{7}$$

where $c_{iv}$ represents the probability amplitude of each of the basis states $|i_v\rangle$ in the initial pure state. It was shown in the Appendix of Ref. [6] that the solution to equation (6) can be written in a factored form at all subsequent time:

$$\rho_{iv,jv'}(t) = \alpha_{iv}(t)\alpha^*_{jv'}(t) \tag{8}$$

where the $\alpha_{iv}(t)$ are complex coefficients given by the solution to the set of equations

$$\frac{d\alpha_{iv}(t)}{dt} = \frac{1}{i\hbar}\sum_j \hat{V}_{iv,jv}\alpha_{jv}(t) - \frac{1}{2}\Gamma_i\alpha_{iv}(t) \qquad (i=1,3). \tag{9}$$

We will assume that the interaction is sufficiently weak that perturbation techniques can be used. In that case and to lowest order, it is apparent from Eqs. (3) and (9) that the total probability amplitude $A_1$ that the atom will absorb two photons and make a transition to the second excited state will correspond to a coherent superposition of the probability amplitudes corresponding to different values of $v$ in Eq. (3). Thus it will be convenient to first solve for the probability amplitude $\alpha_1$ that the atom will be in the state $|h\rangle$ after the absorption of two photons with well-defined frequencies corresponding to a specific value of $v$ with $\omega_i = \omega_{02} - v$ and $\omega_s = \omega_{01} + v$. Subsequently, we will integrate the result over a Gaussian distribution of all possible two-photon frequencies that sum to the energy of the atomic transition $|g\rangle \to |h\rangle$, as described in Eq. (3).

If the interaction is so weak that the initial state is not depleted significantly, the population of the ground state can be taken to be unity. In addition, the intermediate state will not be significantly depleted by transitions to the upper level. In that case, the components of Eq. (9) can be reduced to

$$\dot{\alpha}_1(t) = \frac{1}{i\hbar} m_2^* e^{i\delta t} \alpha_2(t) - \frac{1}{2}\Gamma_2 \alpha_1(t),$$
$$\dot{\alpha}_2(t) = \frac{1}{i\hbar} m_1^* e^{-i\delta t} \alpha_3(t) - \frac{1}{2}\Gamma_1 \alpha_2(t), \qquad (10)$$
$$\alpha_3(t) \equiv 1$$

with the initial conditions $\alpha_1(0) = \alpha_2(0) = 0$. The relevant atomic matrix elements are given by

$$m_1^* = \langle -\mathbf{d}_1 \cdot \mathbf{E}_s(\vec{r}) \rangle = -e^{i\beta(\omega_s)z} d_1 |\mathbf{E}_s(r,\varphi)|,$$
$$m_2^* = \langle -\mathbf{d}_2 \cdot \mathbf{E}_i(\vec{r}) \rangle = -e^{i\beta(\omega_i)z} d_2 |\mathbf{E}_i(r,\varphi)| \qquad (11)$$

Here $\mathbf{E}_s(\vec{r})$ is the classical electric field mode with angular frequency $\omega_s$ and unit polarization vector $\boldsymbol{\varepsilon}_s$ and $\mathbf{E}_i(\vec{r})$ is the classical electric field mode with angular frequency $\omega_i$ and unit polarization vector $\boldsymbol{\varepsilon}_i$. The dipole moments $d_1$ and $d_2$ represent an average over random atomic orientations and the detuning is now $\delta = \omega_s - \omega_1$.

Using the methods of Ref. [6], the probability amplitude $\alpha_1$ that the atom will be in the state $|h\rangle$ in the steady-state limit is

$$\alpha_1(\vec{r}) = \frac{-4i e^{i(\beta(\omega_s)+\beta(\omega_i))z} d_1 d_2 |\mathbf{E}_s(r,\varphi)||\mathbf{E}_i(r,\varphi)|}{\hbar^2 (2\delta + i\Gamma_1)\Gamma_2}. \qquad (12)$$

We now integrate this result using a Gaussian distribution for $\nu$ as in Eq. (2) in order to get the total probability amplitude $A_1$ that the atom will be in the second excited state:

$$A_1(\vec{r}) = -4i d_1 d_2 \sum_\nu N_\nu e^{-\nu^2/(2\sigma^2)} \frac{e^{i(\beta(\omega_{s0}+\nu)+\beta(\omega_{i0}-\nu))z} |\mathbf{E}_s(r,\varphi)||\mathbf{E}_i(r,\varphi)|}{\hbar^2 (2\Delta + 2\nu + i\Gamma_1)\Gamma_2}. \qquad (13)$$

Here $\Delta = \omega_{s0} - \omega_1$ is the mean detuning in the intermediate state. For sufficiently small bandwidths of the down-converted photons, dispersion can be neglected and the propagation constants can then be expanded to lowest order in $\nu$ as

$$\beta(\omega_{s0}+\nu) \approx \beta(\omega_{s0})+\frac{d\beta}{d\omega}\nu,$$
$$\beta(\omega_{i0}-\nu) \approx \beta(\omega_{i0})-\frac{d\beta}{d\omega}\nu. \tag{14}$$

Inserting Eq. (14) into Eq. (13), we have

$$A_1(\vec{r}) = -4ie^{i(\beta(\omega_{s0})+\beta(\omega_{i0}))z}d_1d_2\sum_\nu N_\nu e^{-\nu^2/(2\sigma^2)}\frac{|\mathbf{E}_s(r,\varphi)||\mathbf{E}_i(r,\varphi)|}{\hbar^2(2\Delta+2\nu+i\Gamma_1)\Gamma_2} \tag{15}$$

It can be seen that the frequency-entangled nature of the photon pairs allows all of the frequency components to add coherently and on resonance with the upper transition [7-13]. This would not be possible for two classical pulses of light, and this enhances the two-photon absorption rate beyond the corresponding rate.

Eq. (15) was obtained using periodic boundary conditions. It is shown in the Appendix that the total probability amplitude $A_1$ in the continuum limit becomes

$$A_1(\vec{r}) = -ie^{i(\beta(\omega_{s0})+\beta(\omega_{i0}))z-(2\Delta+i\Gamma_1)^2/(8\sigma^2)}$$
$$\times d_1d_2\sqrt{\frac{2\sqrt{\pi}L}{\sigma u}}|\mathbf{E}_{s0}(r,\varphi)||\mathbf{E}_{i0}(r,\varphi)|\frac{Erfi\left(\frac{2\Delta+i\Gamma_1}{2\sqrt{2}\sigma}\right)-i}{\hbar^2\Gamma_2}. \tag{16}$$

Here $u = d\omega/d\beta$ is the group velocity of the wave-packets, $\mathbf{E}_{s0}$ is the classical electric field mode with angular frequency $\omega_{s0}$, $\mathbf{E}_{i0}$ is the classical electric field mode with angular frequency $\omega_{i0}$, and *Erfi* is the imaginary error function. This result assumes that the bandwidth of the down-converted photons is sufficiently small that the classical field modes do not vary significantly over their frequency range. The two-photon absorption rate in the steady-state limit is then determined by the decay rate out of state $|h\rangle$ and is given by

$$R_2(\vec{r}) = |A_1|^2\Gamma_2 = \frac{2\sqrt{\pi}L}{\sigma u}e^{-(4\Delta^2-\Gamma_1^2)/(4\sigma^2)}$$
$$\times(d_1d_2)^2|\mathbf{E}_{s0}(r,\varphi)\mathbf{E}_{i0}(r,\varphi)|^2\frac{\left|Erfi\left(\frac{2\Delta+i\Gamma_1}{2\sqrt{2}\sigma}\right)-i\right|^2}{\hbar^4\Gamma_2} \tag{17}$$

These results can be integrated over all possible locations of an atom outside of the taper for a given density $\rho_A$ and using the known form of the field modes [14]. This results in a total TPA rate of

$$R_2 = \frac{2\sqrt{\pi}L}{\sigma u} e^{-(4\Delta^2 - \Gamma_1^2)/(4\sigma^2)} (d_1 d_2)^2 \frac{\left|Erfi\left(\frac{2\Delta + i\Gamma_1}{2\sqrt{2}\sigma}\right) - i\right|^2}{\hbar^4 \Gamma_2}$$
$$\times \int_{V_Q'} d^3\vec{r} \left|\mathbf{E}_{s0}(r,\varphi)\mathbf{E}_{i0}(r,\varphi)\right|^2 \rho_A \qquad (18)$$

where $V_Q'$ denotes an integral over all space outside of the core. The integral over the field modes was performed numerically using Mathematica.

## IV. RESULTS

Typical two-photon absorption rates for monochromatic (unentangled) photons in a tapered fiber were estimated in Ref. [6]. In order to compare with that result, we will consider a tapered diameter of 350 nm with a length of 5 mm as before. The central wavelength of the signal and idler photons will also be assumed to be the same as in Ref. [6], where both photons corresponded to a wavelength of 778 nm (degenerate case). We also consider the $5S_{1/2} \to 5P_{3/2} \to 5D_{5/2}$ transition in rubidium with $\Delta = 2.1$ nm, $d_{1,2} = r_{1,2}q$ with $r_1 = 0.223$ nm and $r_2 = 0.0492$ nm, $\Gamma_1 = \Gamma_2 = 10^9 s^{-1}$, and $\rho_A = 10^{12}/cm^3$, as before. The bandwidth of the down-converted photons was assumed to correspond to a wavelength spread of $\sigma = 1$ nm, which is typical of many experiments.

Under these conditions, the TPA rate for frequency-entangled photons was calculated to be $1.45 \times 10^6 s^{-1}$. For comparison, the TPA rate for monochromatic photons was $2.7 \times 10^4 s^{-1}$, which is two orders of magnitude smaller than that for the entangled photons under the same conditions.

The enhancement of the TPA rate for frequency-entangled photons is well known [7-13] and can be qualitatively understood as follows. In order to simplify the discussion, assume that $\sigma \ll \Delta$, so that all the contributions from different values of $\nu$ will correspond to approximately the same detuning. In that case, $Erfi(z) \approx \pi^{-1/2}e^{z^2}/z$. Thus the total TPA rate for entangled photons reduces to

$$R_2 \approx \frac{64L}{u\sqrt{\pi}} (d_1 d_2)^2 \frac{\sigma}{\hbar^4 (4\Delta^2 + \Gamma_1^2)\Gamma_2} \int_{V_Q'} d^3\vec{r} \left|\mathbf{E}_{s0}(r,\varphi)\mathbf{E}_{i0}(r,\varphi)\right|^2 \rho_A \qquad (19)$$

Comparing Eq. (19) with the previous results for coherent states at single-photon intensities (Eq. (16) in Ref. [6]), it can be seen that the entanglement enhances the TPA rate by a factor of $L\sigma/(u\sqrt{\pi})$.

This enhancement factor can be further understood by considering the coincidence rate $R_c$ to detect the signal photon at a position $\vec{r}_1 = (r_1, \varphi_1, z_1)$ and the idler photon at a position $\vec{r}_2 = (r_2, \varphi_2, z_2)$. This coincidence rate is proportional to

$$R_c = \langle \psi_0 | \hat{E}_s^{(-)}(\vec{r}_1) \hat{E}_i^{(-)}(\vec{r}_2) \hat{E}_i^{(+)}(\vec{r}_2) \hat{E}_s^{(+)}(\vec{r}_1) | \psi_0 \rangle$$
$$\propto \left| \int d\nu N_{\beta(\omega_{s0}+\nu)} N_{\beta(\omega_{i0}-\nu)} N E_s(\vec{r}_1) E_i(\vec{r}_2) e^{-\nu^2/(2\sigma^2)} \right|^2 \tag{20}$$

Here $N_{\beta(\omega_{s01}+\nu)}$, $N_{\beta(\omega_{i0}-\nu)}$ and $N$ are the appropriate normalization factors for the signal electric field, idler electric field and initial state, respectively. Considering the fact that the electric fields and the normalization factors vary slowly over the effective bandwidth of the Gaussian wave packet, the integration is much more sensitive to the exponential phase. That is,

$$R_c \propto \left| \int d\nu e^{i\beta(\omega_{s0}+\nu)z_1} e^{i\beta(\omega_{i0}-\nu)z_2} e^{-\nu^2/(2\sigma^2)} \right|^2 \propto \left| \int d\nu e^{i\nu(z_1-z_2)/u} e^{-\nu^2/(2\sigma^2)} \right|^2$$
$$= e^{-\sigma^2(z_1-z_2)^2/u^2} \left| \int d\nu e^{-[\nu/(\sqrt{2}\sigma) - i\sigma(z_1-z_2)/(\sqrt{2}u)]^2} \right|^2 = 2\pi\sigma^2 e^{-\sigma^2(z_1-z_2)^2/u^2} \tag{21}$$

where $u = d\omega/d\beta$ is the group velocity of the wave-packet. It can be seen that the typical longitudinal separation $s$ between two of the entangled photons is described by a Gaussian distribution with a half-width $s = u/(\sqrt{2}\sigma)$. This is equivalent to confining the photons into a mode volume with length $s$ rather than L, which enhances the TPA rate by a factor of roughly $L/s = \sqrt{2}L\sigma/u$. Aside from a factor of $\sqrt{2/\pi} \sim 1$, this is the same enhancement factor that was obtained by comparing the results for entangled and unentangled photons using Eq. (19).

Larger TPA rates could be obtained, for example, by reducing the average detuning $\Delta$ from the first atomic transition, as can be seen in Fig. (3a). This assumes that the detuning is larger than the bandwidth $\sigma$, as illustrated in Fig. (4a), so that all of the frequency components of the signal photon experience approximately the same detuning. On the other hand, the bandwidth of the down-converted photons is a crucial parameter for optimizing the TPA rate. It is desirable to have a large bandwidth so that both photons will arrive at an atom at very nearly the same time, which also enhances the TPA rate. But if the bandwidth is too wide there exist two-photon amplitudes that don't leverage the small detuning from the intermediate state. In addition, as the bandwidth is further increased there will be signal photon amplitudes that have the opposite detuning, as illustrated in Fig. (4b). Two signal photon amplitudes with opposite detuning will give contributions to the two-photon absorption amplitude that will tend to cancel out, since the TPA rate is proportional to the inverse of the detuning. The dependence of the two-photon absorption rate on the bandwidth for a fixed detuning of 2.1 nm (or an angular frequency of 6.5 THz) is shown in Fig. (3b), where it can be seen that the TPA rate first increases for increasing bandwidth but then reaches a maximum at a bandwidth of approximately $\sigma = 1.71$ nm (an angular frequency of 5.3 THz).

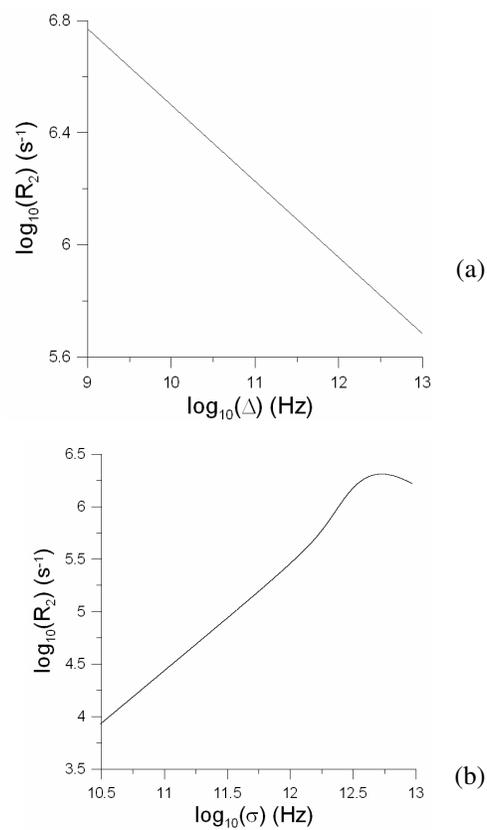

Fig. 3. (a). Dependence of the two-photon absorption rate $R_2$ on the detuning $\Delta$ of the central frequency in the intermediate atomic state, where the bandwidth $\sigma$ is set to 1 nm (or an angular frequency of 3.11 THz). (b). The dependence of two-photon absorption rate on the bandwidth $\sigma$, where the detuning $\Delta$ is set to 2.1 nm (or an angular frequency of 6.54 THz).

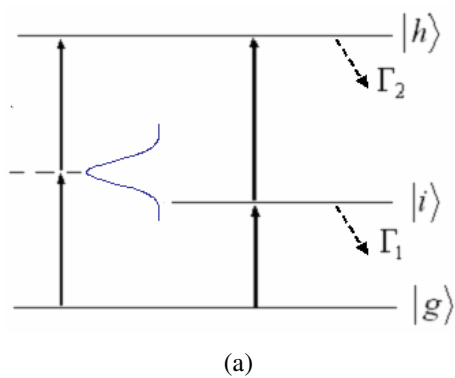

(a)

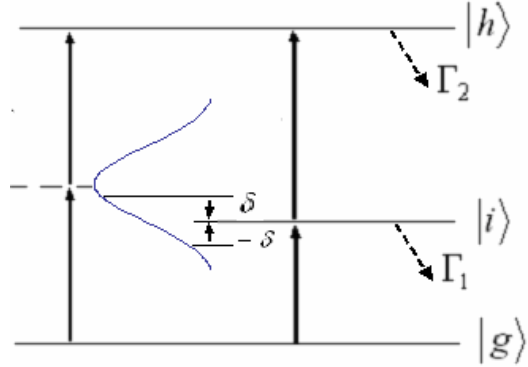

(b)

Fig. 4. Dependence of the TPA rate on the bandwidth $\sigma$ of the photons. (a) When the bandwidth is relatively small compared to the average detuning $\Delta$, most of the frequency components of the signal photon have a detuning with the same sign and contribute constructively to the TPA rate. (b) When $\sigma > \Delta$, some frequency amplitudes correspond to detunings $\delta$ with the opposite sign and their contributions to the probability amplitude for TPA tend to cancel out, since the probability amplitude is inversely proportional to the detuning.

## V. COMPARISON WITH TORIODAL MICROCAVITIES

Toroid microcavities with small mode volumes have been extensively investigated and are expected to have a number of important applications [16-18]. Fig. 5 shows a model of a microtoroid with principle diameter $D$ and minor diameter $d$. In comparison with a tapered fiber, the electric field in the microtoroid is further confined in the longitudinal direction as well as the transverse confinement of a tapered fiber, which leads to a very small mode volume for a single photon. In addition, these devices can have very low intra-cavity losses as demonstrated by reported quality factors greater than $10^8$ [17]. The small mode volume and the ultra-high quality factor make them potential candidates for quantum computing applications. For example, a potential implementation of a Zeno quantum logic gate using resonant cavities of this kind is discussed in Ref. [2], where two tapered fibers are used to couple single photons to two microtoroids.

It is interesting to compare the enhancement in the TPA rate that can be achieved using the small mode volume of a toroidal microcavity and monochromatic (unentangled) photons with that obtained using frequency-entangled photon pairs in a tapered fiber. The argument presented in the previous section suggests that the TPA rates should be comparable if the typical separation $s$ between the frequency-entangled photons is comparable to the circumference of the microtoroid.

The TPA rate in a microtoroid was calculated using methods that are similar to those used in Ref. [6] to calculate the TPA rate in a tapered fiber. A perturbation solution to the electric field of a microtoroid is given in [18]. We considered the case in which the ratio $d/R \ll 1$, which simplifies the calculation of the field modes. The circumference of the toroid was chosen to be equal to s for comparison with the entangled photon case ($D = s/\pi \approx 19\ \mu m$). The minor diameter $d$ was chosen to be 350 nm, which is the same diameter assumed for the tapered fibers; most toroidal microcavities have larger values of d, but this also allows direct comparison with the tapered fiber results. We assumed two counter-propagating monochromatic photons with the same wavelength (778 nm), which corresponds once again to the Rb $5S_{1/2} \to 5P_{3/2} \to 5D_{5/2}$ transition and a relatively large detuning $\Delta = 6.5$ THz. Under these conditions, the TPA rate in the microtoroid was found to be $\sim 0.6 \times 10^6\ s^{-1}$, which is comparable to that obtained for a tapered fiber using frequency-entangled photons. Table 1 compares the rate of TPA for the physical systems that we have considered, assuming the same transverse dimensions.

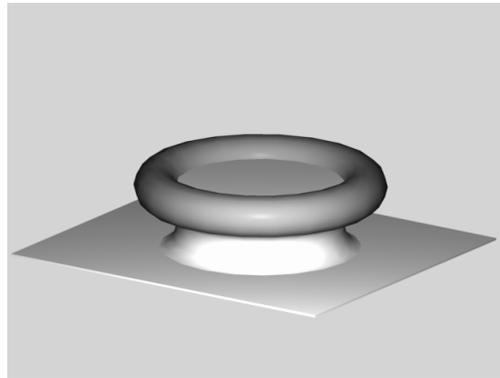

(a)

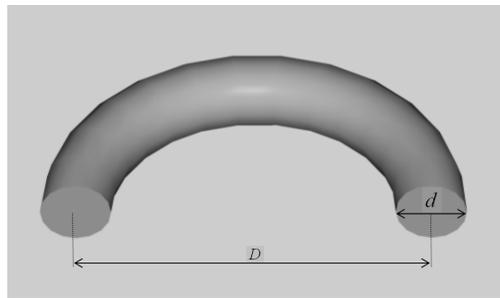

(b)

Fig. 5. (a). A model of a toroidal microcavity fabricated on top of a silicon pedestal. (b). A cut-away drawing of the microtoroid illustrating the principle diameter $D$ and the minor diameter $d$.

| Physical system | Features | TPA rate $(s^{-1})$ |
|---|---|---|
| Tapered fiber and monochromatic photons | Small mode area | $2.7 \times 10^4$ |
| Tapered fiber and frequency-entangled photons | Small mode area and non-classical enhancement | $1.45 \times 10^6$ |
| Microtoroid and monochromatic photons | Small model volume (all three dimensions) and high quality factors | $0.6 \times 10^6$ |

Table 1. Comparison of the TPA rate in several different systems with the same transverse dimensions. These results correspond to photon pairs with degenerate frequencies, while much higher rates can be obtained for smaller detunings. The TPA rate for monochromatic (unentangled) photons in a tapered fiber corresponds to an intensity giving an average of one photon in length L (5mm).

## VI. SUMMARY AND CONCLUSIONS

We have calculated the expected rate of two-photon absorption for frequency-entangled photons in a tapered optical fiber whose diameter is comparable to the wavelength of the light. The confinement of the electric field in the transverse direction increases the intensity associated with a single photon and increases the TPA absorption rate. The TPA rate is further enhanced by the fact that the sum of the frequencies of the two photons is on resonance with the upper atomic state, even though each photon has a relatively broad linewidth [7-13]. This nonclassical enhancement of the TPA rate can be qualitatively understood as being equivalent to confining the photons in the longitudinal direction to a typical separation $s$, which is inversely proportional to their bandwidth $\sigma$. As a result, the photons are effectively confined in all three dimensions, and the TPA rate for frequency-entangled photons in a tapered fiber was found to be comparable to that in a microcavity.

The rate of single photon absorption must be much less than the TPA rate for quantum computing applications. The TPA rates in a tapered fiber are not sufficiently high compared to the single-photon losses to be useful for quantum computing. It may be possible to reduce the single photon loss using electromagnetically induced transparency (EIT) [19], but frequency-entangled photons are not suitable for general-purpose quantum logic operations that must remain valid for arbitrary input qubit states. Nevertheless, these results provide additional insight into two-photon absorption that may eventually be useful in implementing classical logic and memory operations [5] as well as quantum logic devices [1,2].

**ACKNOWLEDGEMENTS**

We would like to acknowledge valuable discussions with T.B. Pittman. This work was supported in part by the Intelligence Advanced Research Projects Activity (IARPA) under United States Army Research Office (USARO) contract W911NF-05-1-0397, and the National Science Foundation (NSF) under grant 0652560.

**APPENDIX**

In the main text, we evaluated the total probability amplitude $A_1$ for TPA absorption (Eq. 15) using periodic boundary conditions. The main purpose of this appendix is to convert to the limit of continuous wavelengths and to perform the necessary integrals to derive Eq. (16).

The normalization factor $N_\nu$ is determined by the requirement that

$$1 = \langle \psi_0 | \psi_0 \rangle = |N_\nu|^2 \langle G_{atom}| \langle 0| \sum_{n,n'=-\infty}^{\infty} \hat{a}(\omega_{i0} - \nu_n) \hat{a}(\omega_{s0} + \nu_{n'}) e^{-\nu_n^2/(2\sigma^2)}$$
$$\times \hat{a}^+(\omega_{s0} + \nu_n) \hat{a}^+(\omega_{i0} - \nu_n) e^{-\nu_n^2/(2\sigma^2)} |0\rangle |G_{atom}\rangle = |N_\nu|^2 \sum_n e^{-\nu_n^2/\sigma^2}. \quad (A1)$$

To evaluate this probability amplitude, the summation over the angular frequency is converted into an integral in the usual way:

$$\sum_n \to \frac{L}{2\pi u} \int_{-\infty}^{\infty} d\nu. \quad (A2)$$

Here $u = d\omega/d\beta$ is the group velocity of the photon wave-packet.
Eq. (A1) then becomes

$$1 = |N_\nu|^2 \sum_n e^{-\nu_n^2/\sigma^2} \to 1 = |N|^2 \frac{L}{2\pi u} \int_{-\infty}^{\infty} e^{-\nu^2/\sigma^2} d\nu = |N|^2 \frac{L}{2\pi u} \sqrt{\pi} \sigma \quad (A3)$$

As a result, the normalization factors is given by $N_\nu \to N = \sqrt{2\sqrt{\pi} u/(L\sigma)}$.

This gives the total probability amplitude $A_1$ in the continuum limit as

$$\begin{aligned}
A_1(\vec{r}) &= -4ie^{i(\beta(\omega_{s0})+\beta(\omega_{i0}))z} d_1 d_2 \sqrt{\frac{2\sqrt{\pi}u}{L\sigma}} \frac{L}{2\pi u} \int_{-\infty}^{\infty} dv \frac{|\mathbf{E}_s(r,\varphi)||\mathbf{E}_i(r,\varphi)|e^{-v^2/(2\sigma^2)}}{\hbar^2(2\Delta + 2v + i\Gamma_1)\Gamma_2} \\
&\approx -ie^{i(\beta(\omega_{s0})+\beta(\omega_{i0}))z} d_1 d_2 \sqrt{\frac{2L}{\pi^{3/2}\sigma u}} |\mathbf{E}_{s0}(r,\varphi)||\mathbf{E}_{i0}(r,\varphi)| \int_{-\infty}^{\infty} dv \frac{e^{-v^2/(2\sigma^2)}}{\hbar^2(\Delta + v + i\Gamma_1/2)\Gamma_2} \quad (A4) \\
&= -ie^{i(\beta(\omega_{s0})+\beta(\omega_{i0}))z - (2\Delta+i\Gamma_1)^2/(8\sigma^2)} d_1 d_2 \sqrt{\frac{2\sqrt{\pi}L}{\sigma u}} |\mathbf{E}_{s0}(r,\varphi)||\mathbf{E}_{i0}(r,\varphi)| \frac{Erfi\left(\frac{2\Delta + i\Gamma_1}{2\sqrt{2}\sigma}\right) - i}{\hbar^2 \Gamma_2}
\end{aligned}$$

where we have used the fact that the electric field varies slowly over the effective bandwidth of the Gaussian wave packet so that $\mathbf{E}_s(r,\varphi) \approx \mathbf{E}_{s0}(r,\varphi)$ and $\mathbf{E}_i(r,\varphi) \approx \mathbf{E}_{i0}(r,\varphi)$. The function $Erfi[z]$ corresponds to the imaginary error function $erf[iz]/i$.